# Analysis of Galaxy Morphology and Evolution using the Pointwise Dimension

Jason S. Best[1], Jane C. Charlton[1,2], and Gottfried Mayer-Kress[3]

## ABSTRACT

In this paper we explore the application of the pointwise dimension (PD) analysis as a large-scale structure descriptor to the RC3 catalog of galaxies (de Vaucouleurs et al. 1991). This technique, which originated in the field of fractal geometry (Mandelbrot 1983) and found many applications in non-linear dynamics, is particularly illuminating in the study of correlations between morphology and environment. To our knowledge the technique is being applied for the first time to the study of galaxy morphology and evolution. The PD is the slope of the curve representing the number of galaxies within some radius, determined separately for each galaxy in the catalog. Thus the distributions of PD's can be compared for subsamples based on morphological type or luminosity. The conclusions of this analysis of RC3 are: 1) The PD elucidates the tendency for early-type galaxies to cluster more than late-type galaxies; however, there is considerable overlap between these populations. 2) The PD statistic does not find a significant relationship between luminosity and clustering in RC3, but it could be an effective diagnostic in larger 2D catalogs. 3) The majority of galaxies in RC3 (excluding clusters) are spread out in space much like a random distribution. A population of objects that traces the distribution of galaxies, but avoids clusters, would have only small correlations. It is worthwhile to pursue application of the PD to larger catalogs with a wider range of luminosities.

*Subject headings:* Cosmology: Large-scale Structure of Universe — Galaxies: Clustering — Methods: Analytical

---

[1]Astronomy and Astrophysics Department, Pennsylvania State University, University Park, PA 16802
[2]Center for Gravitational Physics and Geometry, Pennsylvania State University
[3]Center for Complex Systems Research, Beckman Institute, Physics Dept. UIUC, Urbana, IL 61801

## 1. Introduction

For some time it has been known that the morphological fractions (of ellipticals, lenticulars, and spirals) are related to the density of the environment (Hubble & Humason 1931). The fraction of ellipticals ranges from less than 10% in the lowest density environments to more than 50% in the centers of galaxy clusters (Dressler 1980). It is still not apparent whether this is a result of conditions conducive to formation of a particular morphological type (nature) or to an evolutionary process (nurture).

One general type of nature scenario (Zurek, Quinn & Salmon 1988) requires that the star formation rate depends on the level of the density fluctuation that forms a galaxy. An elliptical galaxy results from a large fluctuation in which stars deplete the gas before it can form a disk. Nurture scenarios have been devised to evolve in the direction from spiral to elliptical as well as in the opposite direction. A spiral or lenticular galaxy could be stripped of its disk by tidal or pressure forces in the center of a galaxy cluster, thus losing its gas and becoming an elliptical (Gunn & Gott 1972). Alternatively, since ellipticals are traditionally thought to be old galaxies, it may be that they form first and somehow acquire disks at a later stage. Finally, scenarios for the origin of ellipticals and lenticulars from mergers have been quite successful (Toomre 1977, Hernquist 1993). Since mergers are known to occur, as evidenced by systems such as the Antennae and "Atoms for Peace" systems (e.g. Barnes & Hernquist 1992), this interactive nurture mechanism surely plays some role in development of morphology.

It is very difficult to determine which of the various nature and nurture mechanisms dominate. It would appear that the clustering properties of different morphological types should be diagnostic of the formation mechanism, but in practice the distinctions are very subtle. If nature is the motivating factor then whatever local properties are needed to make ellipticals could tend to exist over a large region. If a nurture mechanism involving mergers is the dominant mechanism, then initial correlations in the positions of spirals could lead to numerous mergers in one region. Thus in both mechanisms, a correlation between morphology and environment will arise.

Whitmore et al. (1993) argues for a hybrid model of origin of morphology in which the same fractions of spirals, lenticulars, and ellipticals are set by nature to form everywhere in the Universe. The process of formation of a particular type will begin equally often anywhere. However, in this model the process of spiral formation is not as likely to reach completion in the centers of galaxy clusters, and additional evolutionary processes (such as tidal stripping) also alter the populations. It follows from these ideas that the population of elliptical galaxies outside of galaxy clusters should not have particularly different distributions of galaxies around them than other morphological types that are also not in clusters.

The two-point correlation function is the most common way of describing the distribution of galaxies. As defined by Peebles (1980), the two-point spatial correlation function $\xi(r_{12})$ can be written as

$$\delta P = n^2 \delta V_1 \delta V_2 \left[1 + \xi(r_{12})\right] \quad (1)$$

where $n$ is the galaxy number density, $\delta P$ is the joint probability of finding an object in both volumes $\delta V_1$ and $\delta V_2$ at separation $r_{12}$. It is convenient to compute $\xi$ as

$$\xi(r_{12}) = (N_{real}/N_{rand}) - 1 \quad (2)$$

where $N_{real}$ is the number of galaxy pairs in the observed sample with separations between $r$ and $r + \Delta r$, and $N_{rand}$ is the number of pairs in the same interval for the same number of galaxies distributed randomly over an identical area.

On small scales where the peculiar or orbital velocity dominates the Hubble velocity, or for galaxy catalogs with unknown redshifts, it is useful to define an analogous angular correlation function $w(\theta)$ as

$$w(\theta) = (N_{real}/N_{rand}) - 1 \quad (3)$$

with $N$ representing the number of pairs in the actual or random catalogs with separations between $\theta$ and $\theta + \delta\theta$.

Observational determinations of $\xi(r)$ and $w(\theta)$ yield relationships of a power law form:

$$w(\theta) = A\theta^{-\beta} \quad (4)$$

$$\xi(r) = Br^{-\gamma} \quad (5)$$

where $\beta = \gamma - 1$ and $\gamma$ is generally in the range $1.7 - 1.8$ for large redshift catalogs (e.g. Davis & Peebles 1983). More specifically, $w(\theta)$ has been determines separately for samples of galaxies of early and late-types in the region of the Pisces-Perseus supercluster (Haynes & Giovanelli 1988). This study clearly demonstrates the greater clustering of earlier types, with $\beta = -0.90$ for early-types and $\beta = -0.65$ for late-types.

The two-point correlation function, however, has some limitations in providing a complete description of the galaxy distribution. Some problems can be solved by studying the higher order moments (Fry et al. 1993, Matsubara & Suto 1994), but these are often difficult to compute. Another problem is that the technique assumes that the galaxy distribution becomes spatially homogeneous on a length scale that is smaller than that of the catalog being analyzed. If this assumption is invalid, the amplitude of the correlation function and the length scale where the function becomes small are sample dependent (Pietronero 1987). The two-point correlation function is also limited in that it averages together amplitudes on a given scale that come from galaxy pairs in many different environments. This can yield a large amplitude for $w(\theta)$ because of a large number of close neighbors for a small fraction of the galaxies in the sample or because of a smaller number of neighbors for nearly all galaxies in the sample. In other words, while the two-point correlation function indicates that the galaxies are correlated, it does not say how many galaxies contribute to give the value. Finally, this statistic is most often applied either to all galaxies or to galaxies of the same morphology (e.g. elliptical-elliptical pairs). Some environmental influences on morphology might have more to do with the total number of galaxies in the environment of a particular type.

This paper develops a technique to describe the distribution of galaxies that originated in the field of non-linear dynamics. This technique, called the pointwise dimension (PD), allows a separate analysis of the neighborhood of each galaxy in a catalog. It is very simple to apply, and can be utilized to describe both two and three dimensional catalogs. In section 2 of this paper we will give the formal definition of the pointwise dimension (PD) and apply it to the description of the distribution of galaxies. As an example of the use of this technique, we have analyzed the RC3 catalog (de Vaucouleurs et al. 1991), which is described in section 3. The results of this analysis are presented in section 4. The RC3 catalog has velocities for most galaxies, and thus allows a comparison of results when redshift information is used and when it is not used. This comparison indicates how we could interpret an application of this technique to a larger, but two-dimensional catalog. The analysis without velocity information is given in section 4.1 and section 4.2 uses the velocities to compute distances so that the PD can be determined on the same physical scale for all galaxies. In section 4.3, we compute the PD of a galaxy considering only other galaxies within 1000 km/s in radial velocity so that they are more likely to be physically associated. Section 4.4 discusses the use of the PD analyses to separately describe the distribution of galaxies on different scales. Section 4.5 addresses the issue of whether there are differences in clustering around galaxies of various luminosities within a morphological type.

Some questions that we have tried to address in this paper are: 1) Are the environments of all elliptical galaxies special, or do field ellipticals have neighborhoods that resemble those of field spirals? In other words, does the amplitude of the two-point correlation function for ellipticals come from all ellipticals or mostly from the center of clusters?; 2) Does the distribution of galaxies reach homogeneity on the largest scales?; 3) What can be learned about the mechanism of origin for galactic morphology from a statistical analysis of the neighborhoods of galaxies of various types? These issues will be discussed in section 5, with particular attention to the Whitmore et al. model (1993) that suggests that morphology is a result of a combination of nature and nurture (in galaxy clusters).

## 2. Introduction to the Pointwise Dimension

The universe has structure on a large range of scales, from galaxy pairs, to clusters, to superclusters. Also, the two-point correlation function of galaxies has a nearly constant slope over more than two orders of magnitude in distance. A self-similar behavior for galaxy clustering suggests that the concept of fractal geometry may apply (Mandelbrot 1983, Peebles 1993) Several types of fractal analysis have been applied to large-scale structure description, such as wavelet transforms (e.g. Martinez et al. 1993) and percolation statistics (e.g. Klypin & Shandarin 1993). The pointwise dimension is particularly useful for analysis of morphology and environment. It also has the distinction of being conceptually simple and easy to apply to two and three dimensional galaxy catalogs.

We can consider a function $N_{\vec{x}_m}(r)$ which is the count of the number of data points within a distance $r$ from a reference point $\vec{x}_m$. It is found that in a log-log representation that there is a scaling region over which a slope can be defined: within that scaling region, which is bounded by $r_{min}$ and $r_{max}$, this slope $d_{\vec{x}_m}$ is interpreted as the pointwise dimension, and is defined as (e.g. Mayer-Kress 1994)

$$d_{\vec{x}_m} = \frac{log(N_{\vec{x}_m}(r_{max})) - log(N_{\vec{x}_m}(r_{min}))}{log(r_{max}) - log(r_{min})} \quad (6)$$

The motivation of the Mayer-Kress paper is to more accurately reconstruct the dimensions of attractors from time series. It can also be applied to a diverse range of problems in other fields (Mayer-Kress 1989, Zbilut et al. 1989)

The PD technique can easily be applied to a catalog of galaxies. First, consider a two-dimensional catalog where for each galaxy an angular position in the sky is specified. For each galaxy we can plot a curve that gives the number of galaxies within angle $\theta$ of that galaxy. If redshifts are available, the angular separations of galaxies can be converted to a projected separation in units of Mpc. (For conversions we use throughout a Hubble constant of 80 km/s/Mpc.) Figure 1a-c illustrate twenty representative curves, plotted in log-log space, for elliptical, lenticular, and spiral galaxies, respectively.

For each galaxy, a dimension can be calculated by applying a least squares fit to each curve, following the prescription of Holzfuss-Mayer-Kress 1986 (hereafter HMK). The term "pointwise" is used because a slope is calculated separately for each curve, i.e. around each point in the distribution. A sample or subsample of galaxies can then be described by the distribution of slopes, and such distributions can be compared between subsamples (e.g. for a given morphology, absolute luminosity, or region of space).

Of course a single slope fit to a curve is not a complete description of the distribution of galaxies around the galaxy represented by the curve. Although the single slope is indicative of the type of environment in which the galaxy resides, two differently shaped curves can certainly lead to the sample least squares fit. However, we can adapt the PD description to separately address structure on various scales, simply by limiting the distance range over which the slope is determined. For example, to address the question of whether the distribution of galaxies becomes homogeneous on the largest scales, we can determine a slope for each galaxy by just fitting over a range of large distances.

There are several reasons why the PD analysis is advantageous: 1). Other correlation descriptors (such as the Grassberger- Procaccia dimension (Grassberger & Procaccia 1983) were criticized by HMK because they effectively average together curves with different amplitudes. This is the case with the two-point correlation function as well, and this is reflected in its inability to distinguish how many galaxies contribute to large correlations observed on small scales. 2). The PD is very simple to compute and understand, partly because it is not normalized by a random catalog. Even for distant galaxies in a magnitude limited catalog, the slope will not be biased by selection effects, since galaxies will be missing roughly equally at all separations. 3). Morphology-environment studies are facilitated by use of the PD. The two-point correlation function has often been used to explore clustering within a morphological type (using only early or late-type galaxies). For some purposes, however, we would like to analyze all galaxies in the neighborhood of a particular galaxy, regardless of morphology. All of the mass in the vicinity of a galaxy may be influential in a determining the properties with which it forms.

## 3. Description of the RC3 Catalog and Selection Criteria

We have chosen the RC3 catalog (de Vaucouleurs et al. 1991) as an example application of the PD description. The total number of galaxies in the catalog is 23,024, and it is complete for galaxies of apparent diameters greater than 1 arcmin, B-band magnitudes brighter than 15.5, and radial velocities less than 15,000 km/s. For this analysis, we considered only the complete catalog, where each galaxy must have either an apparent diameter of greater than 1 arcmin, or a B magnitude brighter than 15.5. This reduces the number to just under 22,000. The analysis must be applied to a continuous region of the catalog, and thus we analyze the northern and southern hemispheres as separate populations. Boundaries of ±20 degrees Galactic latitude are also used in order to avoid the dust lane and an area of undersampling in the catalog. There remain 10493 galaxies in the Northern sample and 8560 in the Southern sample. For 10% of the galaxies radial velocities are not available. These galaxies are also eliminated in any analyses for which velocities are required. Maps of the sample galaxy distribution are given in Figures 2 a-d for all morphologies, and separately for ellipticals, lenticulars, and spirals.

Depending on the question we are addressing, we may specify the morphological type, luminosity range, and radial velocity range of galaxies for which we will determine a PD. The galaxies that satisfy these specifications (and are sufficiently distant from a catalog boundary) are called "primary galaxies". The PD curves are drawn for each primary, including as "secondaries" any other galaxies within a specified projected distance (determined at the velocity of the primary) or angular separation $\Upsilon$. In order to qualify as a primary, a galaxy must be at least this specified distance or angular separation $\Upsilon$ from any catalog boundary. Figure 1 illustrates PD curves for some representative primary galaxies that were required to be at least 10 degrees from the catalog boundary.

## 4. Results of a Pointwise Dimension Analysis of RC3

Interpretations of the slopes that result from this PD analysis are facilitated by considering the expectation for a random distribution. The number of galaxies within a projected separation should depend on the square of the separation, and thus the slope of the typical PD curve will be 2.0. In general, slopes smaller than 2.0 indicate that the clustering is strong on small scales. A slope larger than 2.0 results when the number of neighbors increases very rapidly at large scales (and there are no nearby neighbors).

We can consider several familiar types of galaxy environments and the PD that galaxies in these environments would tend to produce. 1) A galaxy in the center of a cluster would have many nearby neighbors so the PD curve would rise sharply on scales of Mpcs, flatten out beyond the cluster scale, and then steepen again on scales of 10's of Mpcs. 2) A galaxy on the edge of a cluster would have the sharpest rise at the projected distance of the cluster core, but would still flatten beyond the cluster and rise to approach slope 2.0 on largest scales. 3) A galaxy in a small group will have a rise at scales less than 1 Mpc then a flattening and a gradual rise as the scale increases. 4) A galaxy in a void will have a large PD (possibly greater than 2.0) because the curve will rapidly increase beyond the scale of the void boundary.

In part a of this section, we describe the analysis of RC3 as it would be done if we had no radial velocity information. This type of analysis could be performed on any large two-dimensional catalog. However, since we do have velocity information for RC3 we include this in the analysis in part b. This will yield more detailed information about the distribution of galaxies and will also allow a comparison with how much can be determined from an analysis of a two-dimensional catalog.

### 4.1. Analysis of the Two Dimensional Catalog

For the purpose of this subsection, we use only the angular separations and morphologies of the galaxies, and disregard the radial velocity differences. Of course most galaxies included in the PD curves are not at a radial distance near that of the primary, but it is still possible to detect clustering (as with the angular two-point correlation function).

We begin by considering the PD determined over the angular scale of 0 - 10 degrees (for all primary galaxies at least this distance from the catalog edge). The number of galaxies of each morphology, the median of the PD, and the median of the absolute deviations (MAD) are given in Table 1. The MAD quantifies the spread of values by taking the median of the absolute value of the deviations from the determined median of the sample. The PD increases from early to late-types, but there is considerable overlap between the distributions of slopes.

We repeat the same analysis for galaxies drawn from a random catalog. The same numbers of galaxies are randomly designated as spirals, lenticulars, and ellipticals. The distributions of PD's are illustrated in Fig. 3 (along with those for the RC3 analysis). The median values are very close to 2.0 for all three samples in a random distribution, but there is still a substantial spread. The Median and MAD for these distributions are also listed in Table 1. It can be seen that the random catalog has many galaxies which have the same PD's as galaxies in the real catalog. The overlap with the spiral galaxies distribution is particularly evident, though there are clearly some spirals in environments that yield a smaller PD.

### 4.2. Analysis Including Redshifts

Since the radial velocities in the RC3 catalog range up to $\sim$ 20,000 km/s (with 77 percent of all galaxies having determined redshifts less than 8000 km/s) an angular separation of 10 degrees corresponds to different physical scales for the different primary galaxies. At the median radial velocity of the catalog (5084.5 km/s) 10 degrees corresponds to 11.1 Mpcs. Thus we first use the radial velocities to determine the PD curves for primaries over the projected separation range of 0 - 10 Mpc. Fig. 4 shows the results of this analysis, and median values of the PD are listed (separately for Northern and Southern regions) in Table 2. In both the North and Southern regions of the catalog, there is an increase in the median slopes from early-type to late- type galaxies. The medians in Table 2 are somewhat smaller than the values of the PD determined over the angular scale 0 - 10 degrees (Table 1). The smaller values result from nearby galaxies for which 10 Mpc probes smaller scales than does 10 degrees. This is particularly relevant for early-type galaxies for which the small scales are more likely to yield small slopes.

Although we have used the radial velocity information to compare the environments of various galaxies on the same physical scales, we are still considering secondaries at very different redshifts from the primary galaxy. This clearly dilutes the correlations, but we can quantify this effect by applying a restriction in radial velocity difference to the secondary. Since a typical velocity dispersion in a galaxy cluster is 1000 km/s (Dressler & Shectman 1988), we compute the PD over the range 0 - 10 Mpcs using only secondaries within ± 1000km/s of the primary. These results are also listed in Table 3, and the distributions are given in Fig. 5. Clearly, the PD is considerably decreased when galaxies that are obviously not associated with a primary are eliminated. However, a comparison between morphological types still yields the same qualitative result.

### 4.3. Analysis in Velocity Bins

In order to examine the variations in different regions of the catalog, we separately consider primaries in various ranges in radial velocity. Table 4 shows the median PD results from such an analysis for the sets of three morphological types with primaries in the ranges 0-2000, 2000-4000, 4000-6000, and 6000-8000 km/s, while histograms of the distributions of PD's for the four different velocity bins are given in Fig. 6a-d. The catalog is rather sparse beyond the latter bin.

In all cases, the PD is determined using only secondaries that differ from the primary by less than 1000 km/s in radial velocity. It can be seen that in all four of the velocity bins, the value for the median slope is greatest for the spiral galaxies, although the elliptical galaxies do not always have a smaller median than the lenticulars. The smallest PD's result from the 4000 - 6000 km/s bin. This is largely due to the presence of the Coma cluster in the North.

### 4.4. Pointwise Dimensions on Various Scales

To address the issue of structure on different scales, and in particular the question of whether a catalog becomes homogeneous on large scales, the PD can be computed over various ranges of projected separations. We consider four 5 Mpc shells, centered on the primary galaxy. Table 5 presents the median values of the PD determined for the three morphological classes plotted vs. the shell in which the slope was determined. The table gives values without restricting the velocity of the secondary.

The median values of the PD are smaller when determined on the smaller scale of 0 - 5 Mpc. However, if no velocity criterion is placed on secondaries, the median slope does not vary much with distance from primary beyond 5 Mpc (although it differs between early and late-type galaxies). For spiral galaxies the median PD is as large as 1.88 in the 15 - 20 Mpc range.

However, we have seen in previous sections that the correlations are diluted when no velocity criterion is applied. Table 5 further illustrates this point by the decreases in slopes in all separation bins for all morphological types. The change in PD with projected separation, however, is still not very significant beyond 5 Mpcs. In the 15-20 Mpc bin, we can see that the distribution is still not homogeneous by noting that the median slope is still only around 1.7 for spiral galaxies.

It is interesting to note the behavior of the PD for particular classes of primary galaxies, such as those in the 4000 - 6000 km/s radial velocity bin in the Northern region. These are also listed in Table 5. The lenticular galaxies shows a slope that increases, decreases, and then increases again. This indicates that these galaxies are in a particular type of environment, with not too many neighbors in the 5-10 Mpc bin.

The fact that there are certain classes of galaxies with particular values of the slope suggests that a peak in the PD distribution around a particular value (for any subsample) could be a consequence of the individual group of galaxies within a certain restricted region of the catalog. The members of such a group might have similar PD's and could produce a substantial peak. This would not be a generic property of the Universe as a whole if a given slope does not arise from numerous clumps spread through the region being analyzed. This question was addressed by an analysis of scatter diagrams of the difference in PD between pairs of primaries vs. the difference in separation between them. Galaxies separated by less than a couple Mpc always tend to produce a similar PD as we would expect. On larger scales (e.g. at 20 Mpc) we have many pairs that have similar PD's and many that have substantially different PD's. Thus a certain slope can arise from numerous places in the catalog.

### 4.5. Clustering of Galaxies with Respect to Luminosity

In Iovino et al. (1993), the authors claim that "The result of our analysis is, then, that morphology and luminosity are two independent parameters in the determination of the clustering properties of galaxies...". These results are generated in part by using a two-point correlation function. We endeavored to ascertain if the PD could be used to detect this behavior.

In order to study this phenomenon, we measured the PD for each galaxy within six particular linear distances around each galaxy: 1, 2, 3, 4, 5, and 6 Mpc (converting angular separation to physical distance). It is possible to do this for each galaxy for which we have the redshift information. We have also assumed a Hubble constant of 80 km/s/Mpc, as opposed to the value of 100 km/s/Mpc used in Iovino et al. For this analysis, we have divided the galaxies into four velocity bins of 2000 km/s, for all galaxies with velocities between 1 and 8000 km/s.

We should expect that there will be some galaxies for which the PD cannot be determined, since there can be areas of the catalog that have a dearth of galaxies (e.g. voids). The reason that the PD would not be applicable on these small scales is as follows: a log N-log $\theta$ plot can only be fit over the region where points exist on the graph. If a galaxy does not have a neighbor up through a certain distance (namely the distance we want to measure out to), a log-log plot cannot be generated, and hence no slope can be measure out to that distance.

We can compare the luminosities of the galaxies that have a determined PD to those that do not, to examine whether or not the two distributions of galaxies are significantly different. Applying a K-S test in each velocity bin to the two distributions, we cannot conclusively detect a significant difference in the luminosities between the two sets of galaxies. To the depth of this catalog, the voids are not significantly populated by an underluminous class of galaxies. This by no means rules out the view of void populations of dwarf galaxies since the catalog does not include them.

To further address the issue of a luminosity-density relationship, we plotted the galaxies that had a determined PD slope on a slope-absolute magnitude graph, and applied the Spearman Rank-Order Correlation test. In essence, this test can be used to detect correlations in a bivariate sample such as this. Spearman's $\rho$ is defined as

$$\rho = 1 - [6D/(N^3 - N)] \quad (7)$$

where

$$D = \sum_{i=1}^{N} (R_i - S_i)^2 \quad (8)$$

D is the summation of the ranks of the two variables, which in this case are the PD value and the absolute magnitude. A large value of $\rho$ (e.g. 0.95 or greater) indicates a strong correlation or anticorrelation (depending on the sign of $\rho$). We detect little correlation between slope and absolute magnitude for any of the morphological types, in the velocity ranges between 0 and 8000 km/s out to 1, 2, 3, 4, 5, and 6 Mpc respectively. The Spearman test showed in all cases a $\rho$ value between ± 0.35. Thus, there is no detected correlation between the absolute luminosity and the PD value (which is an indication of the local density)

### 5. Discussion and Conclusion

This analysis of two and three dimensional versions of the RC3 catalog has yielded several insights into the distributions of galaxies of various morphological types. This has been achieved by an application of the pointwise dimension (PD) technique. This simple technique involves a determination for each galaxy of the slope of a curve which represents the number of galaxies within a separation vs. the separation. We have made comparisons of the distributions of slopes for various subsamples of morphology, absolute luminosity, and radial velocity.

First, we shall address the first of the three questions posed in the introduction: to what extent the correlations in the galaxy distribution are produced by several galaxies rather than by a large fraction of them. Fig. 3 compares the RC3 analysis results for the PD calculated on the 0 - 10 degree scale with the expected distributions for a random catalog. Even a random catalog has a considerable spread around the median value of 2.0. Thus we see that many galaxies of all morphological types have a distribution of galaxies around them that is similar to members of a random distribution.

Fig. 5 is a similar plot for the RC3 galaxies, but with the slope determined on the scale 0 - 10 Mpc, and with a restriction on the velocities of secondaries (within ± 1000 km/s of the primary). Although the distributions of the PD clearly do differ between the three morphological populations (ellipticals, lenticulars, and spirals) this plot also shows a substantial overlap. Thus some elliptical galaxies live in environments that are similar to the environments of spirals, and some spirals have a similar distribution of neighbors as typical ellipticals.

The point that many galaxies do not seem to have environments that differ from a random distribution is best illustrated by a comparison of Figs. 7 a and b. These figures are plots of the positions of galaxies, in the RC3 and random catalogs respectively, for galaxies divided into four quartiles by the values of the PD's, determined on the 0 - 10 Mpc scale. We can see from Fig. 7a that most of the smallest slopes come from galaxies that are considerably clustered. For instance, note strains of the Coma cluster at the northernmost part of the first quartile of Figure 7a. However, even the random catalog has some modest sized clusters in the lowest quartile. Furthermore, the distributions of galaxies in the quartiles with the largest slopes are virtually indistinguishable between the RC3 and random catalogs. The second quartile of RC3 appears very similar to the third quartile for the random catalog.

The second question posed in the introduction dealt with whether the galaxy distribution is homogeneous on large scales. The PD increases toward 2 as the scale on which it is determined increases, but it never reaches 2. The RC3 catalog is not large enough to adequately address this question (since there are not many galaxies more than 20 Mpc from any edge). We can again emphasize that although the median value does not reach 2, many of the galaxies in the catalog do have PD's that large, when determined on large scales.

Before addressing the implications of this analysis on the origin of galactic morphology we should consider whether it is practical to apply the PD analysis to a two-dimensional catalog. We have seen in sections 4 a-c that the PD increases as we limit to galaxies that are more likely to by physically associated with the primary. However, even an analysis of a two-dimensional catalog yields the same qualitative conclusions about the distribution of slopes for early and late-type galaxies. Although the analysis described in section 4.5 did not find a relationship between luminosity and PD within a morphological class, the PD analysis may be very useful to discover such a correlation in a larger two-dimensional catalog that has a larger luminosity range.

We have also performed a comparison of the PD results to the results of an application of the two-point correlation function description to the same RC3 catalog. In addition to using the full catalogs, we have computed the PD by restricting the morphology of the secondaries as well as the primaries. This allows a comparison to the two-point correlation function determined in a pure sample of a single morphological type. These results are given in Table 6 for a PD determined on an angular scale of 0 - 10 degrees. The analysis of RC3 yields similar values of the slope of the angular two-point correlation function, $\beta$, as the Haynes & Giovanelli (1988) analysis of the Pisces-Perseus region. If the PD is determined purely for neighbors within the same class we expect the values of the PD will change more for the ellipticals and lenticulars than for the spirals as compared to the values in Table 1.

The third question posed in the introduction was whether an analysis of the morphological environment could clarify to what extend morphology is determined by nature and/or by nurture. If the formation of elliptical galaxies requires a particular condition, such as a large surrounding density of other galaxies, then we might expect some remnant of this condition in their present surroundings. This expectation could also hold if the conditions needed for a particular morphology in a nature scenario were correlated over a large scale.

This question is clearly very complex and there is no unique solution. However, this PD analysis of the RC3 catalog has indicated considerable overlap between the neighborhoods of spirals and ellipticals. Fig. 7 a and b show that the primary reason that early and late-type galaxies are commonly said to be clus-tered differently is because of the contribution of early type galaxies in clusters to the two-point correlation function. This is the expectation that we would have from a model in which the primary nurture mechanism takes place only in galaxy clusters. Outside of clusters there is not much difference between the environments around galaxies of various morphologies. This suggests that a model such as Whitmore et al.'s (1993) blend of nature and nurture could be applicable. In this model the same fractions of all morphologies were set to form everywhere, but tidal forces provide a nurture mechanism that can lead to small PD values in clusters. In fact, the relative elliptical population in each of the lower three quartiles ranges between 4% - 8 %, which can be taken as supporting Whitmore et al.'s claim.

In conclusion, we have found that the pointwise dimension is a viable statistical tool for describing the distribution of galaxies. Our application of the PD to the RC3 catalog has illustrated its value in quantification of the morphology-density relation. Application of this simple and practical technique to larger catalogs could elucidate basic properties of the distribution of galaxies on various length scales with respect to their morphologies and luminosities, and thus could direct the construction of theories of galaxy formation.

We would like to thank B. Whitmore, O. Lahav, P. Laguna, E. Feigelson, S. Odewahn, L. Neuschaefer, and P. Anninos for helpful comments and insightful questions and suggestions during the preparation and presentation of this work. One of us (J. Best) was supported by a NASA Space Grant Fellowship and a CIC Predoctoral Fellowship. This work was supported in part by NASA grant NAGW-3571.

**Figure Captions**

Figure 1: 20 representative curves (each representing a single galaxy) generated by the pointwise dimension (PD) analysis of the RC3 catalog for (a) elliptical galaxies, (b) lenticular galaxies, and (c) spiral galaxies. The curve was generate out to 10 degrees for each galaxy.

Figure 2: Plots of the positions of galaxies from the RC3 catalog. The plots contain (a) all galaxies in our full sample, (b) all ellipticals, (c) all lenticulars, and (d) all spirals. The coordinates range from -180°<l<180° and -90°<b<90° for all plots.

Figure 3: Histograms of the normalized distribution of PD values for all galaxies in the RC3 catalog catalog and for a random catalog. The PD was calculated out to 10 degrees for each galaxy, independent of its velocity. Northern and Southern galaxies were combined in all histograms; the x-axis is binned by intervals of 0.2 in the PD.

Figure 4: Histograms of the normalized distribution of PD values for all galaxies in the RC3 catalog with known radial velocities. The PD was determined on the scale 0-10 Mpc for each galaxy. All other galaxies were considered secondaries, regardless of velocity. Northern and southern galaxies were combined in all histograms; the x-axis is binned in intervals of 0.2.

Figure 5: Histograms of the normalized distribution of PD values determined on the 0-10 Mpc scale for all galaxies with known velocities in the RC3 catalog. This graph differs from Figure 4 in that only galaxies within ± 1000 km/s were considered secondaries when generating the PD. Northern and Southern galaxies were combined in all histograms; the x-axis is binned in 0.2 intervals.

Figure 6: Histograms of the normalized distribution of PD values calculated out to 10 Mpc for each galaxy, computed separately for galaxies with velocities in four bins: (a) 0-2000 km/s, (b) 2000-4000 km/s, (c) 4000-6000 km/s, (d) 6000-8000 km/s. Only galaxies within the particular velocity bin were considered secondaries when generating the PD. Northern and Southern galaxies are combined in all histograms; the x-axis is binned in intervals of 0.2. The solid line represents ellipticals, the dotted line represents lenticulars, and the dashed line represents spirals.

Figure 7: Quartile plots, based on PD value, for the galaxies in (a) the RC3 catalog, and (b) a random catalog. The PD was calculated over the range from 0- 10 degrees for each galaxy. Upper panels represent the quartile with the smallest PD value (i.e. most clustered) and the lowest panels contain galaxies with the largest PD values.

### Table 1: PD Median Values over 0-10 Degrees

**RC3 Catalog**

| Galaxy Type | Number of Galaxies | Median value | MAD |
|---:|:---:|:---:|:---:|
| Northern Ellipticals | 614 | 1.58 | 0.23 |
| Northern Lenticulars | 1217 | 1.67 | 0.23 |
| Northern Spirals | 5813 | 1.81 | 0.22 |
| Southern Ellipticals | 349 | 1.63 | 0.25 |
| Southern Lenticulars | 1260 | 1.68 | 0.24 |
| Southern Spirals | 4447 | 1.79 | 0.22 |

**Random Catalog**

| Galaxy Type | Number of Galaxies | Median value | MAD |
|---:|:---:|:---:|:---:|
| Northern Ellipticals | 575 | 1.93 | 0.12 |
| Northern Lenticulars | 1126 | 1.93 | 0.11 |
| Northern Spirals | 5263 | 1.92 | 0.12 |
| Southern Ellipticals | 344 | 1.91 | 0.12 |
| Southern Lenticulars | 1193 | 1.92 | 0.13 |
| Southern Spirals | 4360 | 1.91 | 0.13 |

### Table 2: PD Median Values over 0 - 10 Mpc

| Galaxy Type | Number of Galaxies | Median value | MAD |
|---:|:---:|:---:|:---:|
| Northern Ellipticals | 398 | 1.47 | 0.30 |
| Northern Lenticulars | 698 | 1.56 | 0.28 |
| Northern Spirals | 3716 | 1.78 | 0.25 |
| Southern Ellipticals | 245 | 1.48 | 0.35 |
| Southern Lenticulars | 729 | 1.61 | 0.28 |
| Southern Spirals | 3482 | 1.78 | 0.25 |

Table 3: PD Median Values over 0 - 10 Mpc

Secondaries within ± 1000 km/s of primary

| Galaxy Type | Number of Galaxies | Median value | MAD |
|---|---|---|---|
| Northern Ellipticals | 374 | 1.10 | 0.31 |
| Northern Lenticulars | 652 | 1.17 | 0.35 |
| Northern Spirals | 3415 | 1.43 | 0.39 |
| Southern Ellipticals | 216 | 1.15 | 0.35 |
| Southern Lenticulars | 646 | 1.43 | 0.39 |
| Southern Spirals | 3114 | 1.45 | 0.38 |

Table 4: PD Median Values over 0-10 Mpc in velocity bins

in units of kilometers per second

| Galaxy Type | $0 - 2000$ | $2000 - 4000$ | $4000 - 6000$ | $6000 - 8000$ |
|---|---|---|---|---|
| Northern Ellipticals | 1.12±0.16 | 1.35±0.14 | 1.08±0.39 | 0.97±0.30 |
| Northern Lenticulars | 1.33±0.27 | 1.32±0.25 | 1.04±0.35 | 1.17±0.38 |
| Northern Spirals | 1.61±0.30 | 1.53±0.30 | 1.38±0.39 | 1.41±0.46 |
| Southern Ellipticals | 1.31±0.25 | 1.30±0.45 | 1.16±0.29 | 0.85±0.26 |
| Southern Lenticulars | 1.14±0.17 | 1.31±0.38 | 1.37±0.42 | 1.02±0.42 |
| Southern Spirals | 1.50±0.25 | 1.48±0.35 | 1.50±0.34 | 1.38±0.44 |

## Table 5: PD Median Values over 0 - 20 Mpc in 5 Mpc Shells

### No velocity constraints on secondaries

| Galaxy Type | Number | 0-5 | 5-10 | 10-15 | 15-20 |
|---|---|---|---|---|---|
| Northern Ellipticals | 291 | 1.34±0.42 | 1.62±0.47 | 1.71±0.44 | 1.71±0.37 |
| Northern Lenticulars | 510 | 1.36±0.38 | 1.75±0.40 | 1.71±0.32 | 1.78±0.32 |
| Northern Spirals | 2703 | 1.61±0.38 | 1.90±0.33 | 1.87±0.32 | 1.85±0.31 |
| Southern Ellipticals | 159 | 1.10±0.34 | 1.77±0.26 | 1.77±0.25 | 1.80±0.25 |
| Southern Lenticulars | 524 | 1.41±0.43 | 1.80±0.28 | 1.84±0.27 | 1.86±0.26 |
| Southern Spirals | 2447 | 1.61±0.39 | 1.89±0.29 | 1.89±0.27 | 1.88±0.26 |

### Secondaries within ± 1000 km/s of primary

| Galaxy Type | Number | 0-5 | 5-10 | 10-15 | 15-20 |
|---|---|---|---|---|---|
| Northern Ellipticals | 171 | 0.99±0.31 | 1.21±0.53 | 1.18±0.44 | 1.35±0.39 |
| Northern Lenticulars | 348 | 1.07±0.41 | 1.34±0.48 | 1.35±0.43 | 1.44±0.44 |
| Northern Spirals | 1801 | 1.30±0.47 | 1.60±0.51 | 1.52±0.44 | 1.57±0.43 |
| Southern Ellipticals | 85 | 0.93±0.24 | 1.30±0.27 | 1.44±0.24 | 1.39±0.32 |
| Southern Lenticulars | 310 | 1.11±0.46 | 1.54±0.50 | 1.55±0.41 | 1.54±0.40 |
| Southern Spirals | 1519 | 1.35±0.47 | 1.71±0.45 | 1.67±0.43 | 1.68±0.40 |

### Northern Lenticulars with v=4000 km/s - 6000 km/s

| Galaxy Type | Number | 0-5 | 5-10 | 10-15 | 15-20 |
|---|---|---|---|---|---|
| Northern Lenticulars | 97 | 0.93±0.37 | 1.52±0.45 | 1.23±0.47 | 1.46±0.43 |

## Table 6: Comparison of PD and w($\theta$)

### Scale analyzed: 0 - 10 degrees

| Galaxy Type | $\beta$(Giovanelli-Haynes) | $\beta$(RC3) |
|---|---|---|
| Early | -0.90 | -1.00 |
| Late | -0.65 | -0.60 |

### PD Median Values over 0-10 degrees

| Galaxy Type | Number of Galaxies | Median value | MAD |
|---|---|---|---|
| Ell-Ell | 336 | 1.24 | 0.45 |
| Len-Len | 662 | 1.43 | 0.39 |
| Spi-Spi | 3716 | 1.74 | 0.24 |